\documentclass[aip,reprint]{revtex4-1}
\pdfoutput=1 
\usepackage{misccorr}
\usepackage{amsmath}
\usepackage{graphicx}
\usepackage{ifpdf}
\usepackage[colorlinks=true,bookmarks,bookmarksnumbered=true]{hyperref}
\usepackage{graphicx}
\usepackage{bm}

\begin{document}
\title{Mechanical relaxation behavior of polyurethanes reinforced with the in situ-generated sodium silica-polyphosphate nanophase}

\author{V.O. Dupanov}
\affiliation{Institute of Macromolecular Chemistry, Kyiv, Ukraine}

\author{S.M. Ponomarenko}
\affiliation{National Technical University of Ukraine ``Kyiv Polytechnic Institute'', Kyiv, Ukraine}

\begin{abstract}
Further exploration of hybrid organic/inorganic composites (polyurethane based with inorganic material sodium silica polyphosphate) properties with mechanical relaxometer gives ability to analyze microstructure of such materials in terms of chain reptation tubes filler's  fractal aggregates and stress amplification.
\end{abstract}

\keywords{chain reptation tube; fractal aggregates; nanocomposite; network model;  nanoparticle; phase separation; relaxation process; relaxation time; stress amplification}
\pacs{61.66.Hq, 62.20.Fe, 62.40.+i, 65.40.De}

\maketitle

\section{INTRODUCTION}

	Segmented polyurethanes (SPU) is a generic name for polyblock copolymers which are usually obtained by polyaddition reaction between long-chain di- or polyols, and diisocyanates \cite{Saunders}. Due to poor compatibility between polyol and  diisocyanate chain fragments (commonly referred to as soft and stiff segments, respectively), the latter usually self-associate and segregate as nano-size domains from a continuous phase of the former. Depending on the relative content and the intrinsic properties of soft and stiff nano-phases (i.e., molar weight, polarity, chain stiffness, etc.), the mechanical performance of SPU can be readily varied over an unusually broad range, from soft rubbers to hard thermoelastoplastics \cite{Lipatov, Kercha}. The relatively easy synthesis of SPU combined with their unique properties make them attractive for different practical applications; however, their commercial value is severely limited by quite a few deficiencies, the intrinsically high flammability being among the most notorious \cite{Hu}. An apparently straightforward way to minimize these deficiencies by, e.g., compounding of SPU with mineral fillers, proved to be of little feasibility in so far as their other important properties significantly deteriorated.	
	An attractive alternative route for synthesis of SPU with improved performance can be based on sol-gel technology which ensures a homogeneous, in situ generation of inorganic \cite{Kick} nanoparticles throughout a continuous polymer matrix. Using this route to generate a ceramic-type nanophase (silicates, aluminates, phosphates, etc.), a series of SPU-based nanocomposites (SPUN) with significantly improved weather-, radiation- and flame-resistance was developed. 
This paper is aimed at the further characterization of the same samples \cite{Isch} by mechanical relaxation technique.

\section{EXPERIMENTAL}
\subsection{Materials}

As described \cite{Isch}, the organic/inorganic hybrids (segmented polyurethane-based nanocomposites, SPUN) were prepared by reaction between the organic precursor [macrodiisocyanate, MDIC, from propylene oxide glycol oligomer ($MM = 1052$) and 2,4-toluene diisocyanate], and the inorganic precursor (so\-di\-um silica-polyphosphate, SSP) 

Thin films of the SPUN for physical characterization were cast from solutions using methyl cellulose as a surfactant, and dried overnight in a vacuum oven at $150^{\circ}$ C to constant weight. 

\subsection{Experimental method}
The step-wise loading (stre\-ching)/un\-lo\-ading (contraction) cycles were measured (with the estimated mean errors below 2 \%) at room temperature with the mechanical relaxometer on basis of stretching calorimeter described in details \cite{Mirontsov}.
Each specimen was stretched at a constant velocity  $q_{+}$ (10\% of the total specimen length per minute) to a predetermined   $\lambda_i$  , stored at fixed  $\lambda_i$   to the full completion of mechanical relaxations, and thereafter allowed to contract at the same velocity  $q_{-}$   to zero force. The typical difference between fixed extensions in two successive steps, $\Delta\lambda = \lambda_{i+1}-\lambda_i$, varied from several digits in the fourth place to a few digits in the third place.
Prior to each measurement, the previous mechanical history was erased by several successive pre-stretchings to the maximum extension   $\lambda_{lim}$ , free relaxation to zero stress and subsequent storage in an unloaded state overnight. Then the specimen was stretched to a predetermined fixed extension    $\lambda_{f} < \lambda_{lim}$, and the time dependence of the stress   $\sigma$  was monitored at    $\lambda_{f}=const$.  

\section{RESULTS AND DISCUSSION}

\subsection{Main models}
There's a lot of possibilities to use relaxation method.
Usually for un\-cross\-linked polyurethanes the stress-realxation phenomena considered as a set of physical
processes, such as are: breacking of weak chemical linkages (urethane, allophanate, ets.), disrupting of 
hydrogen and other secondary bonding, decreasing the number of free entanglements and 
slippage of trapped entanglements~\cite{Dziera}. Also reported as a result of different experimental
methods that main role in stress-relaxation in polyurethanes
plays soft segments at the temperatures below 373 K and higher then
 $T_g$~\cite{Dziera,Rheo-optical}. Well known that hysteresys phenomena caused by high
stress-orientation mechanism in hard domens (HD) of PU, at large strains HD even can be brocken
, and high mobility of soft segments is a pledge
of fast stress relaxation and shape recovery~\cite{infrared}.
Several popular models are used for explaining of relaxation datas. First was Maxwell's models.
There are different variations, but general is that solid body presented as a set of viscousious and
elastic modes. Amount of different modes depends on morphological models. Fast and slow modes
reffered to different morphological units (e.g. microdomens~\cite{Bartenev}, entropic elasticity
of back bone segments in the filler-reinforced matrix~\cite{fiberreinforced}, ets.)

One of the drawbacks of this models is limited set of relaxation modes and as a result is impossibility
to include all relaxational spectra in mathematical description. 

Recent times reptational theories based on the tube model of de Gennes, Edwards and 
Doi. It was shown in~\cite{unattached}, that mechanism of relaxation in networks
Model of Kolraush-Watts-Williams (KWW) based on a simple termodinamical 
principles leads to a common fractional power law and consider this problem 
via introduction of stretched exponent which should reflect a common shape of relaxation spectra~\cite{simplemodel}.
Main equation should be written as:
\begin{equation}
\sigma(t)=\sigma_{inf}+\sigma _0 e^{-(t/\tau_p)^{1-n}}
\end{equation}
where $n$ -- coupling parameter, $\sigma_{inf}$ -- non-relaxing part of overall modulus,
$\sigma _0$ -- relaxing part, $\tau_p$ -- effective time. Here $n$ represents 
microscopically the strength of coupling between a relaxing species and it's 
surroundings and macroscopically -- the breadth of relaxation. The higher the coupling
strength between primitive spiecies with it's envirement, the closer $n$ to unity. 

Complications appears when nonlinear effects of stretching gonna be included into
discussion. In this case the popular stohastic reptational models are hardly  
 fits experimental datas well. Main distinguishing nonlinear processes during 
 large step-strain experiments assumed as: chain retractions, convective constraint 
 releases (CCR), partial strand extensions (PSE) and nonaffine 
deformations~\cite{nonlinearrelaxation}.

Modification of KWW model concerning nonlinear effects gives a dependense of 
coupling factor $n$ from time as $n(T_f(t))$, where $T_f$ is a fictive 
temperature, which
ascribes volume changing during relaxation as 
$$T_f-T=(V-V_\infty)/V_\infty\Delta a(T),$$ where $\Delta a(T)$ is the difference 
between liquid and glassy state termal expansion coefficients.
Main equations should be written as:
\begin{equation}
\label{1}
\sigma(t)=\sigma_{inf}+\sigma _0 e^{-(t/\tau_p)^{1-n(T_f))}}
\end{equation}

\begin{equation}
\label{2}
\tau_p=[(1-n(T_f))\omega_0^{n(T_f)}\tau_0]^{1/(1-n(T_f))}
\end{equation}

\section{Experiment and discussion}
\subsection{Parameter $k=1-n$}
Experimental data was calculated accordingly to simple KWW model~\cite{simplemodel}.

\begin{figure}[tb]

\includegraphics[width=\columnwidth]{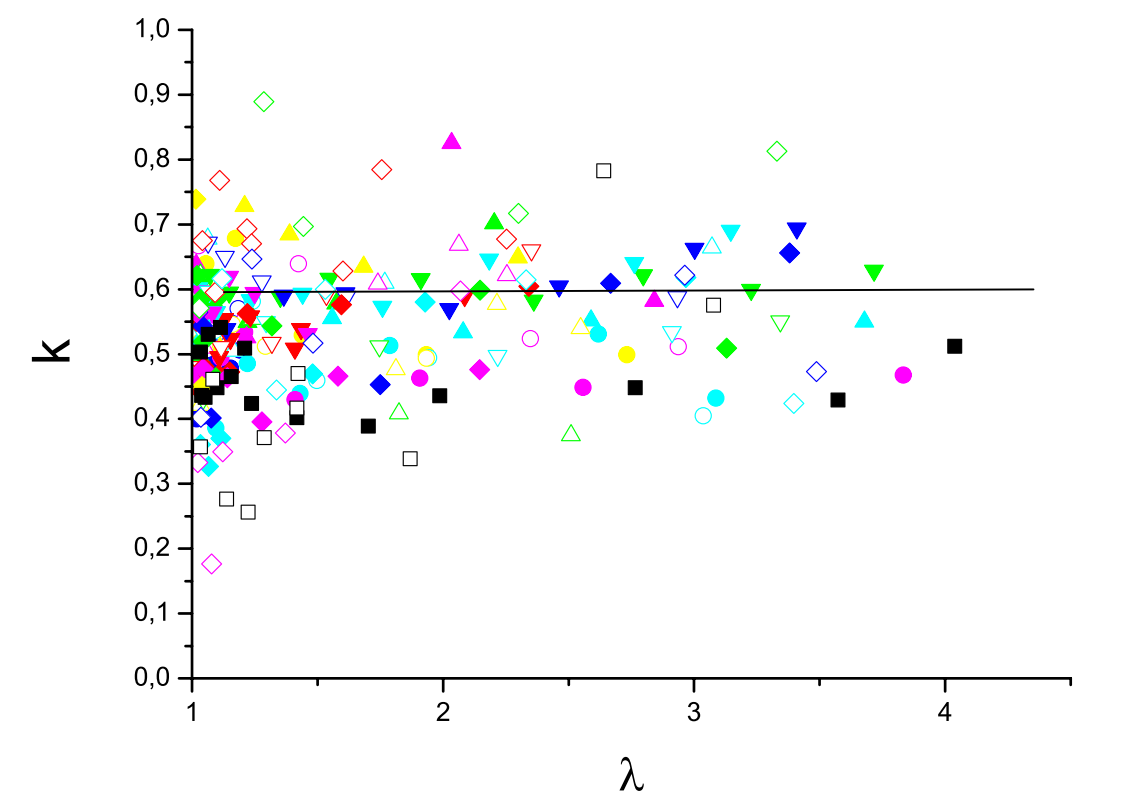}
\caption{\small\label{pic1}Exponental parametr $k=1-n$ of KWW model. Filler content (mas.): 
red symb. -- 10\%, green -- 20\%, blue -- 30\%, cyan -- 40\%, magenta -- 50\%, yellow -- 60\%. Hollow symbols - 
preddeformed samples. Squares -- MDI, circles -- SPh (+ MC), triangles -- SPh (+PMS), upset triang. -- SAPh (+PMS), diamonds -- SAPh (+MC)}
\end{figure}

As can be seen from summary picture for $k$ (Fig.\ref{pic1}), in the wide range of
elongations parameter $k$ does not depend from elongation, which can suggest 
linearity of relaxation process. 
Moreover, the whole set of samples this parameter has approximately the same 
values of $k$: in the range of $0.45 - 0.7$. 
According to this one can say, that $n$ is in the range of $0.3 - 0.55$. The most
 importaint results is the absence of  large divergencies between coupling 
mechanisms in pure PU and filled samples. However $k$ for pure PU slightly
 lower than for filled samples. That could indicate on weak coupling mechanismus
 in filled samples in the reason of domination of phase separation 
above chemical bonding. It means that coupling forses mostly specified by interactions 
of spieces with hard domains even in the case of large filler contents.

\subsection{Stress $\sigma$}

\begin{figure}[tb]
\includegraphics[width=\columnwidth]{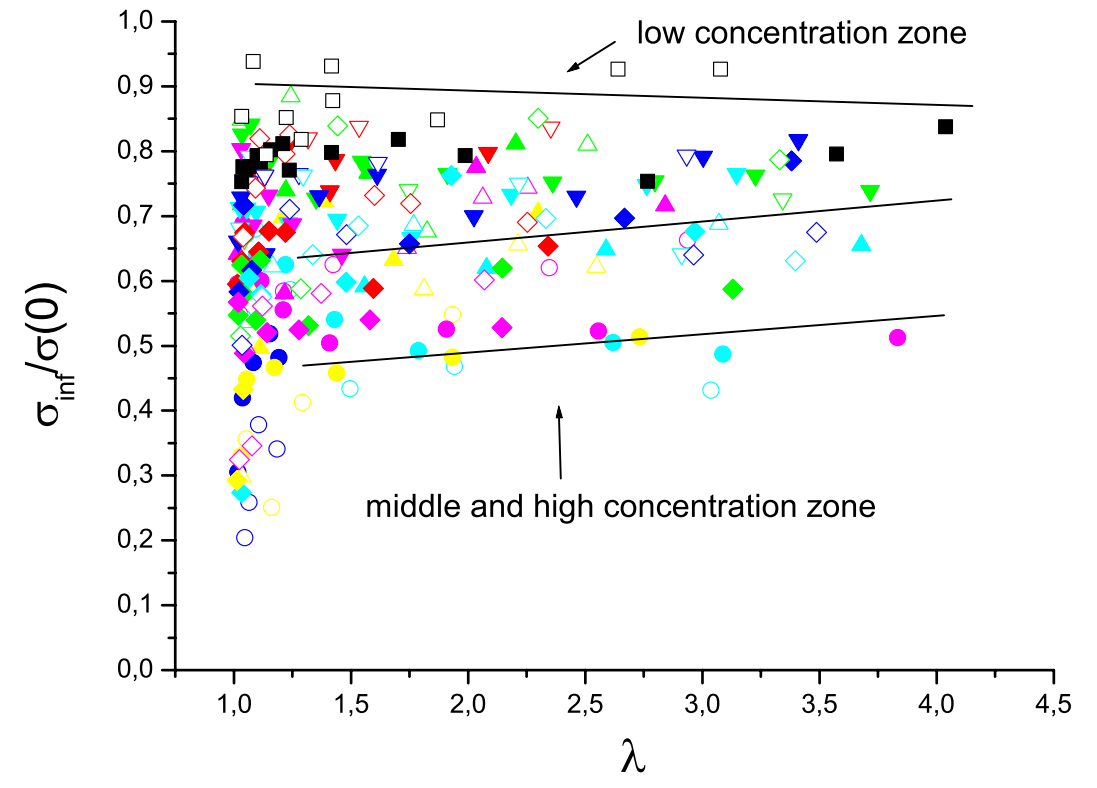}
\caption{\small \label{pic3}Values of $\frac{\sigma_{inf}}{\sigma(0)}$ calculated from KWW model. 
Filler content (mas.): red symb. - 10\%, green - 20\%, blue - 30\%, cyan - 40\%, magenta - 50\%, yellow - 60\%. Hollow symbols - preddeformed samples. Squares - MDI, circles - SPh (+ MC), triangles - SPh (+PMS), upset triang. - SAPh (+PMS), diamonds - SAPh (+MC)}
\end{figure}

The value of redused $\sigma_0$ usually is out of consideration. Main iterest is 
 the behavior of $\sigma_{inf}$. For the most relaxation models this value 
 is out of relaxation spectra dependence and connected with density of crosslinks.
 Classical theries of rubber ellasticity  in the 
case of Maxwell's model for equlibrium modulus gives~\cite{glassfiberreinforced}: 
\begin{equation}
\label{Einf}
E_{inf}=\frac{\sigma_{inf}}{\epsilon}=\frac{3RTd}{M_c}
(1-\frac{2M_c}{M_n})
\end{equation}
where $R$ -- gas constant, $d$ -- density, $T$ -- temperature, 
$M_c$ -- average molecular weight of segment between two crosslinking points, $M_n$ -- 
number average molecular weight of chain.

It could be seen from Fig.~\ref{pic3} that whole group of results can be devided 
onto two filler containt-dependent zones as it shown with straight lines. A 
visiable decrease of equlibrium stress with increasing of filler conteint in the whole 
range of elongations 
indicates on lowering of number of crosslinking points (crosslinking density).
 Comparing this with moduli dependense it could be pointed that decreasing of 
 filler content could leads to lowering of crosslink dencity, but it does not mean 
 lowering of moduli. So inorganic hybryd fillers plays in PU reinforsing role 
 even in the case of \emph{in-situ} reaction process.
Calculation of experimental data in accordance with eq.~\eqref{Einf} gives that 
if $M_n\approx 1000$ than for pure MDI $M_c$ lies in the range from above zero 
to 450, and for 50\% - 60\% wt. filled samples $M_c$ lies from 450 to 765.
\subsection {Parameter $\tau_p$}

\begin{figure}[tb]

\includegraphics[width=\columnwidth]{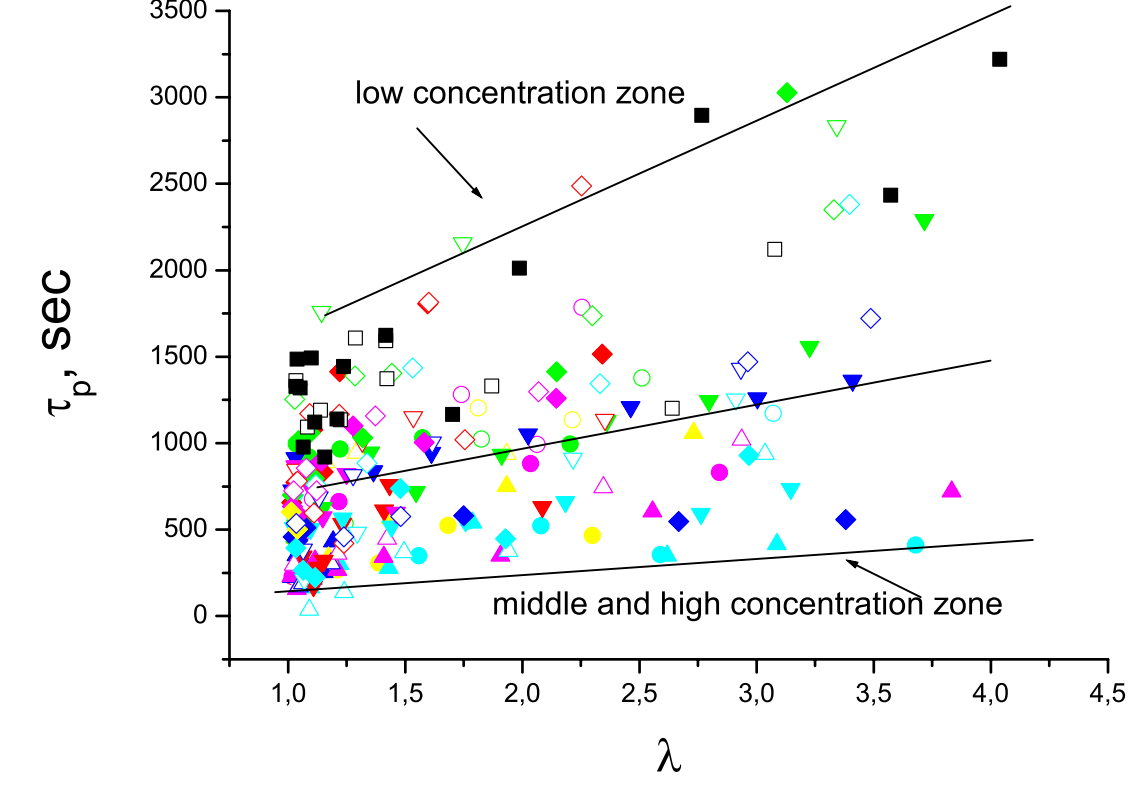}
\caption{\small\label{pic4}Values of $\tau_p$ calculated from KWW model. 
Filler content (mas.): red symb. -- 10\%, green -- 20\%,
blue - 30\%, cyan -- 40\%, magenta - 50\%, yellow -- 60\%. Hollow symbols -- 
preddeformed samples. Squares -- MDI, circles - SPh (+ MC), triangles -- 
SPh (+PMS), upset triang. -- SAPh (+PMS), diamonds -- SAPh (+MC)}
\end{figure}

As can be seen from Fig.~\ref{pic4}, whole set of datas can be divided onto 
two parts the same way as for $\sigma$ datas. Low concentration datas are 
extremely high then datas for highly filled samples.
As shown in \cite{simplemodel}, effective relaxation time and "true" relaxation time
 has found to be identical. As far as $\tau_p$ connected with temperature 
 as Vogel-Fulcher-Tamman-Hesse-like equation:
 \begin{equation}
 \label{3}
 \tau_p\propto exp[B/(T-T_\nu)]
\end{equation}
 and main equation \eqref{2} which determine value of $\tau_p$ is connected both 
with cooperative response time (frequency) and independent relaxation time 
of primitive spieces it could be supposed dependence of relaxation time 
from values of free volume~\cite{polyacetal}. Also it can be related to the 
crosslink dencity~\cite{Dziera}. Interesting interpretation
also gave Irjak~\cite{Irjak} who related relaxation time with concentration 
of equilibrium physical junctions and conformation of chains in the network. 
 Analizing Fig.~\ref{pic4} it can be supposed, that there is a linear dependance 
 between crooslink dencity (density of equlibrium junctions) and stretching ratio.
  \emph{In-situ} generated filler could decrease density of crosslinks and 
  lowers relaxation times. But it can be seen that preddeformation increases 
  relaxation times only in cases of big amount of fillers. May be it was caused 
  by better interaction between already broken clasters of filler with flexible matrix 
  in relation to small filler containing samples.
 
\section{Conclusions}

\begin{enumerate}
\item Shown that even in filled samples coupling mechanismus determinied mostly 
with interactions between hard domens of PU and matrix of PU.
\item There is no or completely small nonlinear effects in the range of elongations up to  300\%.
\item In the case of \emph{in-situ} process of synthesys of hybrid organic/inorganic 
 materials, inorganic filler acts as reinforsing agent for PU.
\item Crosslinking density slightly reduced as filler containt increasing.
\item Preddeformation does not affect relaxation times in case of small filler containt, 
 but increases $\tau_p$ in case of filler containt $\ge40$\%, which can be related 
 to a greater increas of crosslink density after preddeformation.
\end{enumerate}

\section{Additions}
\begin{enumerate}
\item Great plastic deformation occurs in the hard phase of PU and leads to 
hysteresis~\cite{mechpropertiesSPU}.
\item Relaxation mechanism in microblocs with two relaxation processes, 
proposed by Irjak \cite{Irjak} looks like KWW -- Doi, Edwards idea 
with $w_c$ and $w_0$.
\item Stress can be stored mostly in the filler in the reason of increasing 
acting region between phases~\cite{fiberreinforced}.
\end{enumerate}

\end {document}